\documentclass[twocolumn]{svjour3}[2021.3.8]

\usepackage{graphicx}
\usepackage{array}
\usepackage{subfigure}
\usepackage{setspace}
\usepackage{amssymb}
\usepackage{ulem}
\usepackage{booktabs}
\usepackage{stfloats}
\usepackage[justification=centering]{caption}
\usepackage{amsmath}
\usepackage{autobreak}
\usepackage[colorlinks,linkcolor=blue,anchorcolor=blue,citecolor=blue]{hyperref}
\usepackage{algorithm}  
\usepackage{algpseudocode}  
\usepackage{marvosym}
\usepackage{ifsym}

\begin{document}
\title{Hierarchical Deep Network with Uncertainty-aware Semi-supervised Learning for Vessel Segmentation}

\author{Chenxin Li \and Wenao Ma \and Liyan Sun \and \Letter Xinghao Ding \and Yue Huang \and Guisheng Wang \and 
Yizhou Yu
}
\institute{Chenxin Li \and Wenao Ma \and Liyan Sun \and \Letter Xinghao Ding \and Yue Huang \at School of Informatics, Xiamen University, Xiamen 361005, China, \\\email{dxh@xmu.edu.cn}
\and 
Guisheng Wang \at Department of Radiology, the Third Medical Centre, Chinese PLA General Hospital, Beijing, China,\\\email{wanggs1996@tom.com}
\and
Yizhou Yu \at Deepwise AI Laboratory, Beijing 100125, China,\\\email{yizhouy@acm.org}
% \thanks{The work is supported in part by National Key Research and Development Program of China (No. 2019YFC0118100), in part of ZheJiang Province Key Research Development Program (No. 2020C03073), in part by National Natural Science Foundation of China under Grants 81671766, 61971369, U19B2031, U1605252, 61671309, in part by Open Fund of Science and Technology on Automatic Target Recognition Laboratory 6142503190202, in part by Fundamental Research Funds for the Central Universities 20720180059, 20720190116, 20720200003, and in part by Tencent Open Fund.}
}

% \thanks{Chenxin Li, Wenao Ma, Liyan Sun, Xinghao Ding, and Yue Huang are with the
% School of Informatics, Xiamen University, Xiamen 361005, China (e-mail:
% dxh@xmu.edu.cn). }
% \thanks{Guisheng Wang is with the Department of Radiology, the Third Medical Centre, Chinese PLA General Hospital, Beijing, China
% (e-mail: wanggs1996@tom.com)}
% \thanks{Yizhou Yu are with the Deepwise AI Laboratory, Beijing 100125, China
% (e-mail: yizhouy@acm.org)}
% }
% \thanks{y J. Paisley was with Department of Electrical Engineering, Columbia
% University, New York, NY, USA}}% <-this % stops a space

\maketitle
\begin{abstract}
The analysis of organ vessels is essential for computer-aided diagnosis and surgical planning. 
But it is not a easy task since the fine-detailed connected regions of organ vessel bring a lot of ambiguity in vessel segmentation and sub-type recognition, especially for the low-contrast capillary regions. 
Furthermore, recent two-staged approaches would accumulate and even amplify these inaccuracies from the first-stage whole vessel segmentation into the second-stage sub-type vessel pixel-wise classification.
Moreover, the scarcity of manual annotation in organ vessels poses another challenge. 
In this paper, to address the above issues, we propose a hierarchical deep network where an attention mechanism localizes the low-contrast capillary regions guided by the whole vessels, and enhance the spatial activation in those areas for the sub-type vessels. 
In addition, we propose an uncertainty-aware semi-supervised training framework to alleviate the annotation-hungry limitation of deep models. The proposed method achieves the state-of-the-art performance in the benchmarks of both retinal artery/vein segmentation in fundus images and liver portal/hepatic vessel segmentation in CT images.
\end{abstract}
\keywords{Vessel Segmentation \and Hierarchical Deep Network \and Attention Mechanism \and Semi-supervised Learning}

\section{Introduction}
The analysis and visualization of vessels in human organs are essential for diagnosis and treatment of various diseases.
For example, automatic retinal artery and vein (A/V) segmentation assists doctors diagnose systemic and cardiovascular diseases \cite{abramoff_retinal_2010,kawasaki_retinal_2012}. Specifically, the asymmetrical change of retinal A/V measured by arteriolar-venular ratio (AVR), is closely associated with several cardiovascular diseases \cite{ikram_are_2004,wong_retinal_2002}. Decrease in AVR is a biomarker with increased risk of stroke \cite{kawasaki_retinal_2012}. Meanwhile, the narrowing of retinal arteriolar caliber has been reported to be related to the risk of hypertension and diabetes development \cite{chew_retinal_2012,nguyen_retinal_2008}. Similarly, accurate segmentation and visualization of liver vessels are key prerequisites for safe and efficient surgery in the liver regions \cite{SurgLiver1,SurgLiver2}. The analysis of hepatic and portal vessels can also provide valuable diagnostic information with regards to chronic liver disease and cirrhosis \cite{LiverDiag}. 
In a word, these clinical requirements are summarized as the accurate segmentation of organ vessels as well as their sub-types, e.g., the retinal vessels with arteries and veins in fundus images, and the liver vessels with hepatic and portal sub-types in liver CTs.

\begin{figure}[t]
\begin{center}
%\centering
    {\includegraphics[width=8.0cm,height=5.0cm]{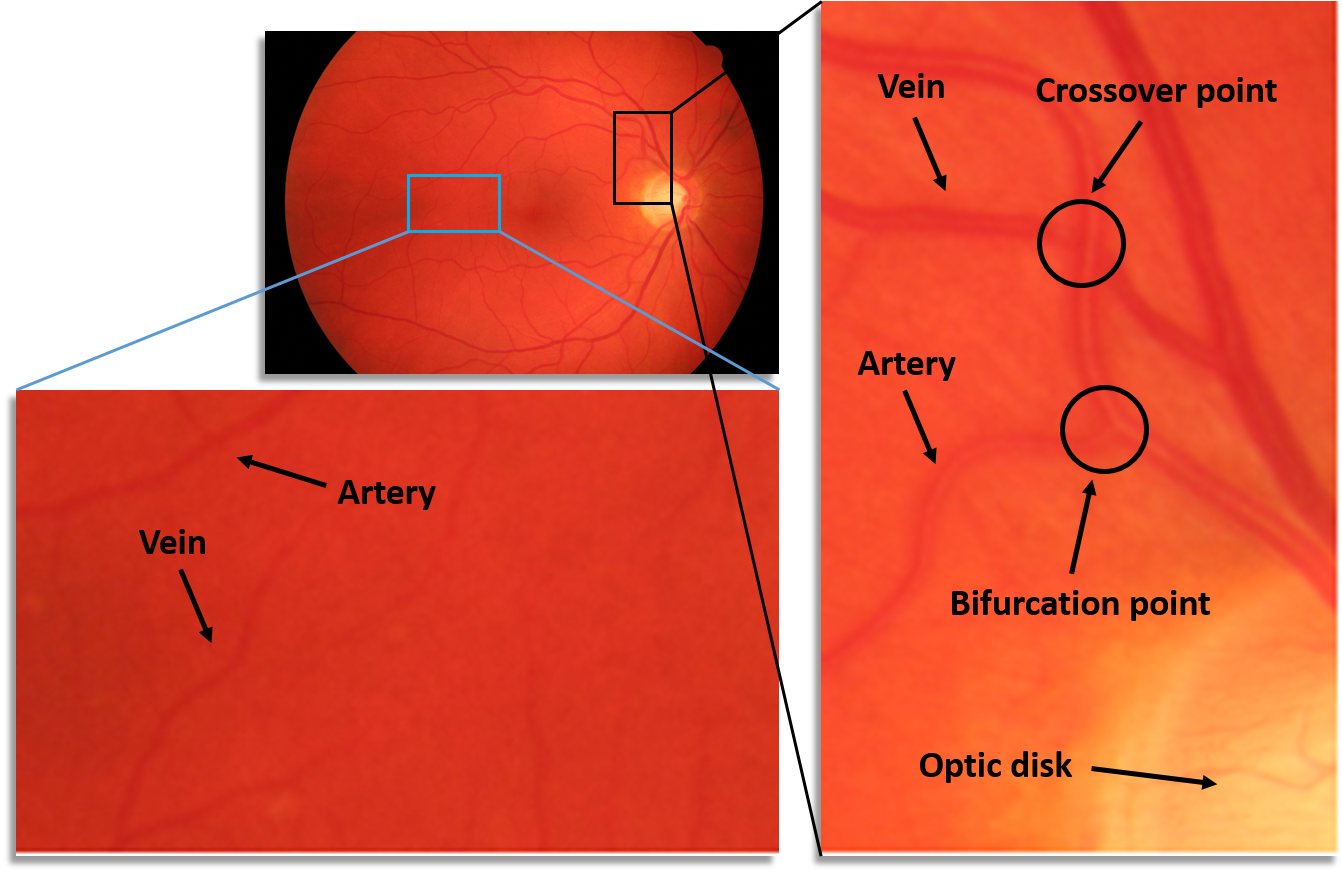}}
    {\includegraphics[width=8.0cm,height=5.0cm]{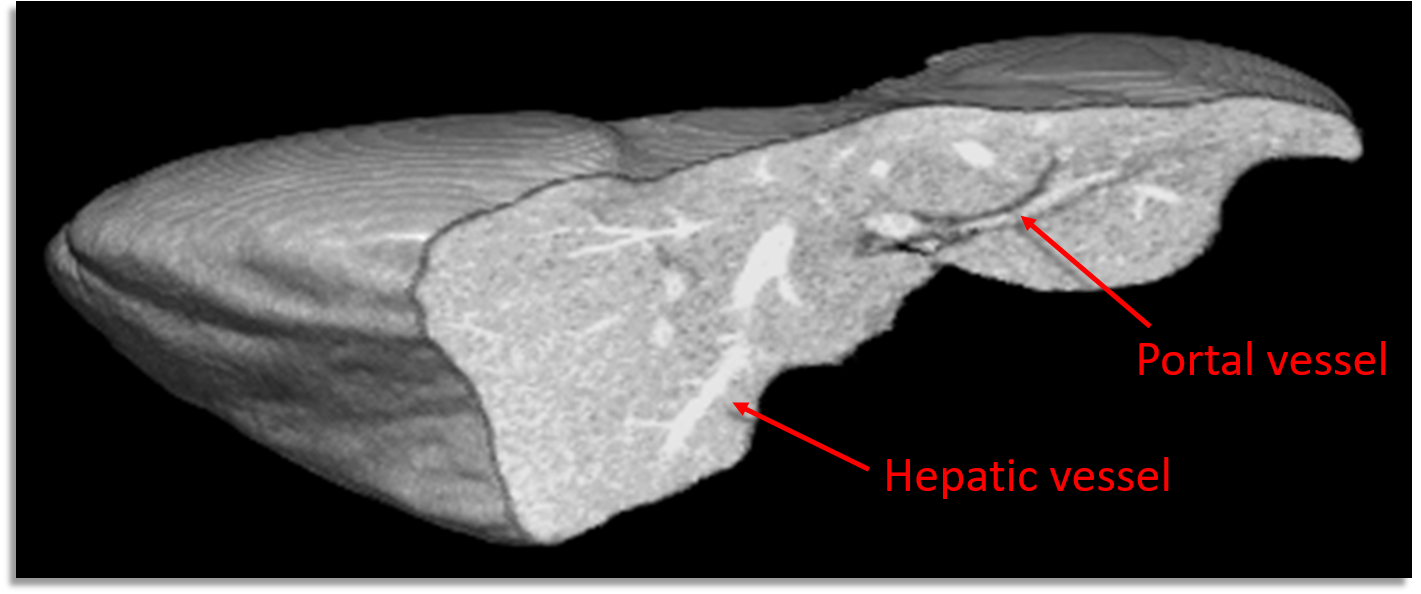}}
    \caption{An example of 3-channel color retinal fundus image and of 1-channel grayscale anatomical CT volume of a liver. In the retinal image, the regions near and far from the optic disk regions are zoomed in for better visualization. In the liver image, the regions of portal vessels and hepatic vessels are marked.} \label{RetinalFundus}
    %\vspace{-0.5cm}
\end{center}
\end{figure}

Although this field has received considerable research attention, several limitations hinder the routine application of vessel segmentation and sub-type classification in the clinical practice. 
First, the low contrast between capillaries and background and between different sub-type vessels poses a challenge in segmentation models.
An example of retinal fundus image is shown in Figure \ref{RetinalFundus}. Some capillary vessels in the areas far away from an optic disk are too indiscernible to be detect and the contrast of A/V is very low there. 
% Furthermore, some capillary vessels are extremely inconspicuous thus difficult to detect. The ambiguity in distinguishing arteries and veins on regions far from optic disk also effect the classification negatively in retinal images.
% We also show an example of anatomical liver CT image in Figure \ref{RetinalFundus}. 
The situation is similar when observing portal and hepatic vessels in liver anatomical CT images, as shown in Figure \ref{RetinalFundus}.
% The low-contrast vessels, especially the venules in liver CT images are difficult to be separated from backgrounds. 
To address this challenge, we think the model should focus more on those low-contrast areas, which are usually capillaries, vessel edges, etc. We discover that the unconfident predictions of whole vessel segmentation are often generated from the low-contrast regions, as shown in Figure \ref{DrivePerformance}. 
With this location information as a guide, we can spatially activate the feature representation of sub-type vessel segmentation, which enhances the spatial attention of the model on those 'hard' regions for the task of sub-type vessel segmentation. 

Besides the above limitation, the previous methods that segment vessels in a two-stage regime, which first segment the whole vessels, then classify the results into sub-types, may propagate and even amplify the first-stage errors mainly caused by the low-contrast regions to the subsequent stage.
Instead, the multi-task fashion has attracted long-standing attention and proved to be effective in other medical practices \cite{vorontsov2018liver,bi2017automatic}. For example, the task of liver lesion segmentation is usually coupled with the liver segmentation in a multi-task manner. Similar in spirit, we integrate the segmentation of whole and sub-type vessels as a one-stage multi-task framework, to improve the feature sharing between the two tasks, alleviate the accumulated errors and further overcome the challenge caused by low-contrast regions.

Another challenge is the scarcity of the pixel-wise annotation for vessels and their sub-types. Usually, the labeling process for pixel-wise vessels requires radiology expertise and is laborious as well as time-consuming, which causes the acquiring labels for every image in a training dataset to be a burden.
Thus, a more practical situation in practical clinical applications is that only some training data are labeled. To address this, we leverage the information from massive unlabeled datasets via semi-supervised learning methods. A common approach is self-labeling, where the model is trained with labeled data by supervised methods and then allocates pseudo labels for unannotated data. However, the pseudo-generated labels are not always reliable due to the distribution shift between labeled and unlabeled data, which initiates the requirement of filtering unreliable pseudo labels. Our intuition is that if 
an unlabeled sample is allocated a more consistent pseudo label, i.e., maintains the same class, we think its pseudo label is high-quality and reliable. Thus,
we introduce the uncertainty estimation for the generated labels, and retain only the ones with enough certainty for the subsequent optimization.

Overall, in this paper, we propose a hierarchical deep network with uncertainty-aware semi-supervised learning for organ vessel segmentation. First of all, we propose the hierarchical capillary-enhanced network for joint whole and sub-type vessel segmentation, where the multi-task design completes both sub-tasks, emphasizes on the capillary-like low-contrast regions based on the guide information from the sub-task of whole vessels segmentation, and regularizes the model to avoid the error propagation occurring in prior two-stage works.
% We couple the segmentation of whole and sub-type vessels in a unified framework, which 
% regularizes the model to avoid the error propagation occurring in two-stage prior works.
% Moreover, based on the observation that capillary vessels are usually allocated predictions with high entropy,  we develop a activation mechanism to push the feature branch of sub-type vessel segmentation task to focus more on these regions.   
% with the guide from the branch of whole vessels. 
% We found vanilla segmentors tend to predict capillary unconfidently with possibilities around $0.5$ where vessels indeed exist, which motivates our attention strategy. 
Moreover, we develop a uncertainty-aware semi-supervised learning method to efficiently leverage information from massive unlabeled datasets. We allow the model to self-label the unlabeled data and filter the low-quality pseudo labels. To further model the reliability of the generated labels, we introduce the uncertainty-aware estimation and filter the pseudo-labeled data with high uncertainty.

% To tackle with these challenges, we proposed a hierarchical capillary-enhanced network for joint whole and sub-type vessel segmentation. The segmentation of whole and sub-type vessels simultaneously utilizes their geometric characteristics and regularize the model to avoid the effect of error propagation occurring in two-stage prior works. Meanwhile, based on the observation on the low-contrast sub-type vessel regions in organs, we designed a spatial activation on sub-type vessels with the aid of attentions from segmentation on whole vessels. We found vanilla segmentors tend to predict capillary unconfidently with possibilities around $0.5$ where vessels indeed exist, which motivates our attention strategy. Also, to efficiently leverage massive unlabeled datasets, we developed a semi-supervised training approach guided by uncertainty modeling. By only leveraging automatically annotated pseudo-labels with high confidence, our model avoids unreliable annotations.

Our initial work has been published on MICCAI 2019 \cite{ma2019multi}. We extend the conference version by incorporating a novel semi-supervised learning framework by uncertainty modeling. We also enrich the experimental section by validating the proposed model on liver vessel segmentation.

% The main contributions of our paper are threefold: first, by adding a spatial activation block, we adopt a multi-task neural network for artery/vein segmentation, which can utilize the result of vessel segmentation to assist artery/vein segmentation; second, we develop a semi-supervised training strategy which enables our model to effectively utilize the unlabeled data, which can help improve the robustness of the model; third, we especially propose a novel approach to achieve uncertainty estimation by combing a conditional variational auto encoder with the segmentation network, which is computationally cheaper than other uncertainty estimation approaches.

\section{Related work}
In this section, the previous models for organ vessel and sub-type vessel segmentation are reviewed. The semi-supervised learning methods that have been employed in the case of shortage of labeled data are then reviewed. 

\subsection{Organ Vessel Segmentation}
In graph-based approaches, a vascular graph is built from the extracted vessel centerlines and then each individual graph tree is classified into arteries and veins. The model proposed by Niemeijer \textit{et al.} \cite{Niemeijer2010} extracts handcrafted features from the vessel centerlines and then performs classification based on soft labels. Dashtbozorg \textit{et al.} \cite{dashtbozorg_automatic_2014} designed a model for retinal A/V segmentation through a pipeline of whole vessel segmentation, centerline extraction and graph reconstruction. Estrada \textit{et al.} \cite{estrada_retinal_2015} proposed a graph-theoretic framework for A/V segmentation by estimating the vascular topology using a global likelihood model and domain specific features. Xu \textit{et al.} \cite{xu_improved_2017} extracted novel texture features from vessel centerlines and employed KNN as the classifier. In the recent work from Zhao \textit{et al.}\cite{zhao_retinal_2018}, the dominant set theory was used to identify the complicated branches and crossover junctions, building a vascular graph and classifying the centerlines into artery and vein. 
In terms of liver vessel segmentation, a contextual information extraction was used in a hierarchical voting mechanism in \cite{livervessel2} to segment liver vessel. 
% Centerlines constraint and intensity model built on kernel fuzzy c-means are integrated to a graph cut \cite{livervessel3}. 
Incorporating knowledge of liver vascular structures, Jerman \textit{et al.} \cite{jerman2016enhancement} employed vascular filters based on Hessian filters with a loss constraint ratio to the vessel shape.
Merveille \textit{et al.} \cite{merveille2017curvilinear} utilized morphological filters by ranking the orientation responses of path operators. Lebre \textit{et al.} \cite{lebre2019automatic} further proposed a 'Couinaud' scheme for the portal/hepatic vessel segmentation. 
Guo \textit{et al.} \cite{guo2020novel} modeled a vascular network based on graph cut, thinning and vessel combination.
Nevertheless, the above two-step methods suffer from the limitation that the performance of sub-type vessel classification affects the accuracy of the whole vessel segmentation, especially for the graph-based methods. Additionally, feature-engineering methods heavily rely on the fine design of hand-crafted feature extraction, which presents difficulty in capturing complex vessel patterns. 

With the rapid development of deep neural networks, the number of studies using Fully Convolutional Networks (FCN) to detect and classify organ vessels is increasing. AlBadawi and Fraz \cite{albadawi_arterioles_2018} adopted the FCN with an encoder-decoder structure for the pixel-wise prediction of retinal arteries and veins. Meyer \textit{et al.}\cite{Meyer2018} also used FCN for the same task and reported the performance on major vessels thicker than three pixels. 
Huang \textit{et al.} \cite{huang2018robust} utilized FCN 3D-UNet for the vessel segmentation in liver CT images.
Deep learning based methods have demonstrated their potential to segment sub-type vessels in an end-to-end manner. However, these models applied direct segmentation without exploiting contextual image information, thus showing uncertainty in prediction and often classifying pixels on sub-type vessels especially in capillary as background. Thus there is still room for improvements if the specific model design could be proposed to account for the ambiguity in prediction and incorporating proper contextual information. 

\subsection{Semi-supervised Learning}
With the aim at exploiting a large number of unlabeled data, plenty of researches raised the focus on semi-supervised learning strategy \cite{hong2015decoupled,hung2018adversarial,li2018weakly-,wei2018revisiting,papandreou2015weakly-and}. Hong \textit{et al.} \cite{hong2015decoupled} proposed decoupled networks including classification and segmentation networks, where bridging layers can deliver class-specific information. Hung \textit{et al.} \cite{hung2018adversarial} utilized an adversarial learning mechanism and adopted a logit (probability maps) output from network as a confidence map in semi-supervised learning. Moreover, some approaches also designed unsupervised loss into the overall loss function for semi-supervised learning \cite{Yves2005Semi,Laine2016Temporal,Tarvainen2017Mean}. Sajjadi \textit{et al.} \cite{Sajjadi2016Mutual} proposed a consistency loss on the outputs of the model conditioned on random perturbations of one image to impose transform invariance. In some prior literature, researchers assigned pseudo-labels to the massive unlabeled datasets \cite{Lee2013icmlw,shi2018transductive}. Lee \textit{et al.} \cite{Lee2013icmlw} utilized a pre-trained network to generate pseudo-labels of unlabeled data by itself, which were then used for finetuning the model to achieve better robustness.

\begin{figure*}[thb]
\centering
\includegraphics[width=15.0cm,height=6cm]{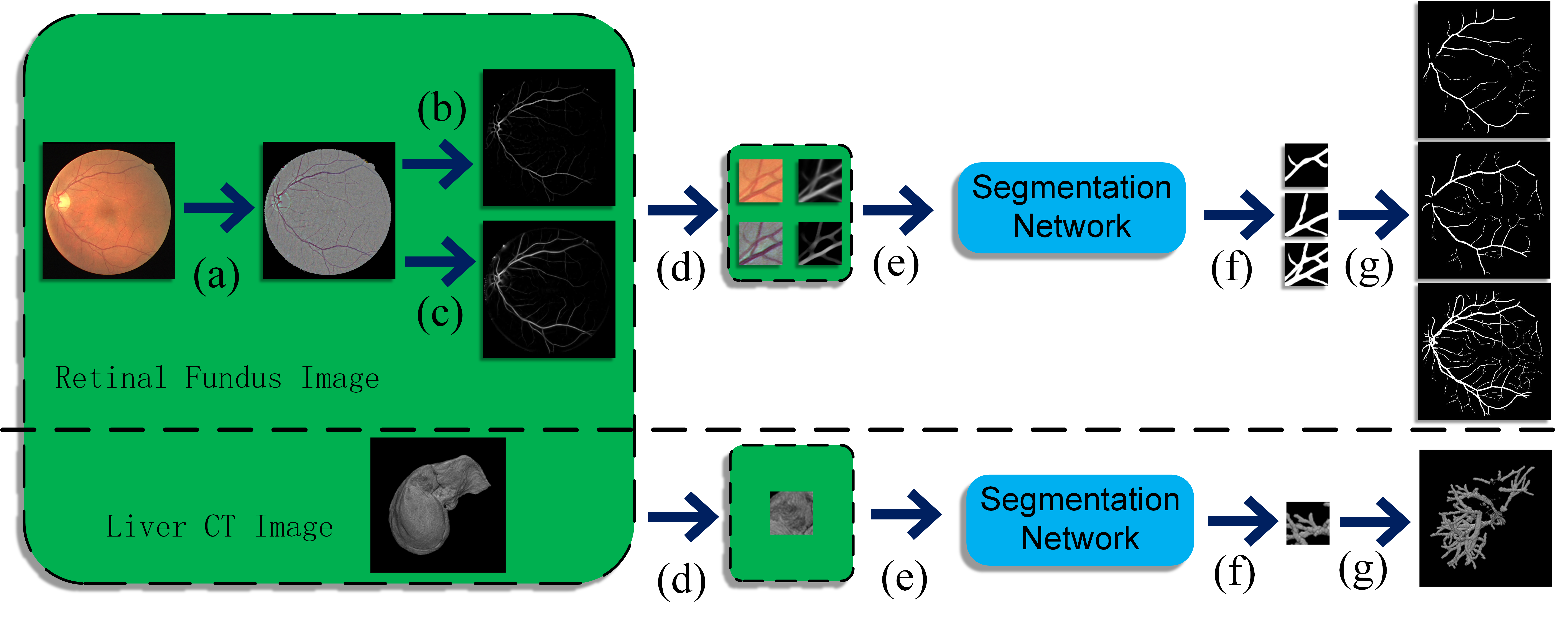}
\caption{The basic pipeline of the segmentation for retinal A/V and liver vessels.  A multi-input module is applied for retinal A/V, which leverages the results from three common image pre-processing techniques \cite{ma2019multi}.
(a) Illumination correction; (b) Gabor filtering; (c) Line detector; (d) Patch extraction; (e) Feeding patches into segmentation network; (f) Logits of patch vessel segmentation; (g) Segmentation on whole image.} \label{Overflow}
%\vspace{-0.5cm}
\end{figure*}

\begin{figure*}[ht]
\centering
\includegraphics[width=17.0cm,height=7.5cm]{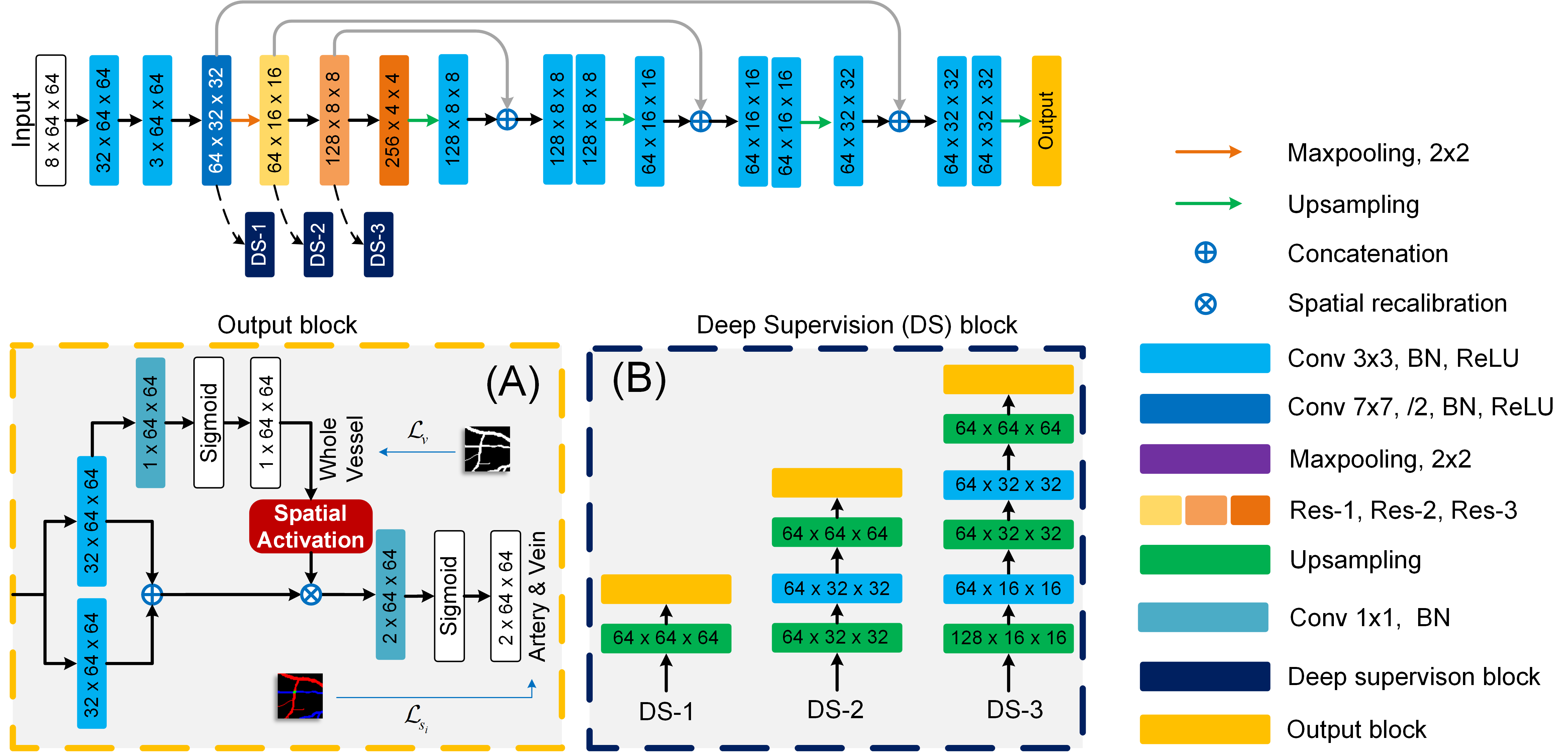}
\caption{The architecture of our proposed hierachical deep network. (A) Output block; (B) Deep supervision branches.} \label{fig3}
%\vspace{-0.5cm}
\end{figure*}

\section{Method}
%The system workflow of our proposed algorithm is shown in Figure~\ref{fig2}. 
A data pre-processing pipeline and a hierachical network are described in this section. Then the design of spatial activation to enhance low responses in sub-type vessels in capillary regions is detailed. Finally, the uncertainty-based semi-supervised framework built on a probabilistic model is presented. Please note that a retinal fundus image is used to illustrate our approach.

\subsection{Data Pre-processing}
We use specific pre-processing on different organs and modalities. For retinal fundus images, we apply a series of pre-processing on raw images shown in the green box in Figure \ref{Overflow}. To correct the bias generated by non-uniform illumination, a illumination correction is applied. Then a line detector \cite{PreProc2} and Gabor filter \cite{PreProc1} are utilized on corrected images in parallel for edge extraction. Raw images, illumination-corrected images and two edge maps are further cropped into pairs of patches and are further concatenated in the channel dimension to form an 8-channel input (i.e., 3 channels for a raw/corrected image, 1 channel for either of the two kinds of edge map). In the ablation study, we validate the effectiveness of this proposed multiple-inputs design. In the testing phase, a sliding window strategy is applied on the test image for the patch-wise predictions, as shown in Figure \ref{Overflow}, then the patches are further merged to reconstruct the image prediction via averaging on logits of overlapping patches as post-processing.

For liver vessel segmentation validated on 3D liver CT scans, we randomly crop volumetric cubes from normalized liver CT scans as model input without other pre-processing operations in the training phase. A similar sliding-window approach is applied in testing phase.

\subsection{Hierachical Capillary-enhanced Network}
Here we present the architecture of our hierarchical capillary-enhanced network in Figure \ref{fig3} for joint whole and sub-type vessel segmentation. A U-Net like architecture is adopted as backbone, since U-Net \cite{UNET} has proven its potency in segmenting medical images accurately and efficiently. Similar to the regular U-Net, we concatenate the feature maps from the previous decoding layer with the corresponding feature maps from the encoding layers to be used as input in each decoding layer.

To predict the whole vessel and sub-type vessels (i.e., arteries/veins in retinal image and portal/hepatic vessels in liver image) simultaneously, we design an ad-hoc output block where contextual information from whole vessel segmentation helps guide the sub-type vessel segmentation, as shown in Figure \ref{fig3} (A) . The motivation of this design is building a hierarchical model to aid a difficult task (sub-structure segmentation) with information from an easy task (whole vessel segmentation) and avoid accumulative segmentation error.

We apply two different sets of convolution on the input features of the output block to generate two chunks of feature maps. For whole vessel segmentation, one chunk of features is squeezed into a 1-channel representation via $1\times1$ convolution + BN, then a Sigmoid function is utilized to produce a logit ranging from 0 to 1, where a binary cross-entropy loss function $\mathcal{L}_v$ is imposed. Therefore, the output block is able to predict whole vessel segmentation while the resulting logit describes the contextual information of the confidence level of the model in making a prediction about a certain pixel belonging to a vessel instead of background. In this stream the common characteristics of sub-type vessels are extracted as the whole vessel regions to be distinguished from backgrounds.

As shown in the output block in Figure \ref{fig3} (A), two chunks of features obtained by convolving input features are concatenated together in another stream and weighted by the spatial activation which is discussed in the next subsection. Then the weighted features are further forwarded into a $1\times1$ convolution + BN + Sigmoid layer, squeezing them into a logit of N channels with N denoting total number of sub-type vessel classes. Each channel in the logit represents the likelihood of this sub-type class being presented in the prediction. In this paper, N equals 2 for both retinal and liver vessel segmentation. Then each channel of the logit is used for computing the binary cross-entropy loss to train the segmentation of sub-type vessels $\mathcal{L}_{s_i}$ where subscription $i$ represents a certain sub-type vessel. Note that the segmentation of sub-type vessels is not a one-hot problem since a pixel is likely to be classified into both artery and vein if it lies in the junction of vessels. Thus either channel in the two-channel logits represents the possibility of one substructure of vessel. In this stream, the model learns to differentiate sub-type vessels (artery/vein in the retinal image or portal/hepatic vessel in the liver image). To designing such a two-stream multi-task output block, we exploit both the homogenous and heterogenous characteristics of all sub-type vessels.

\begin{figure}[t]
	\centering
	\includegraphics[width=8cm,height=5.0cm]{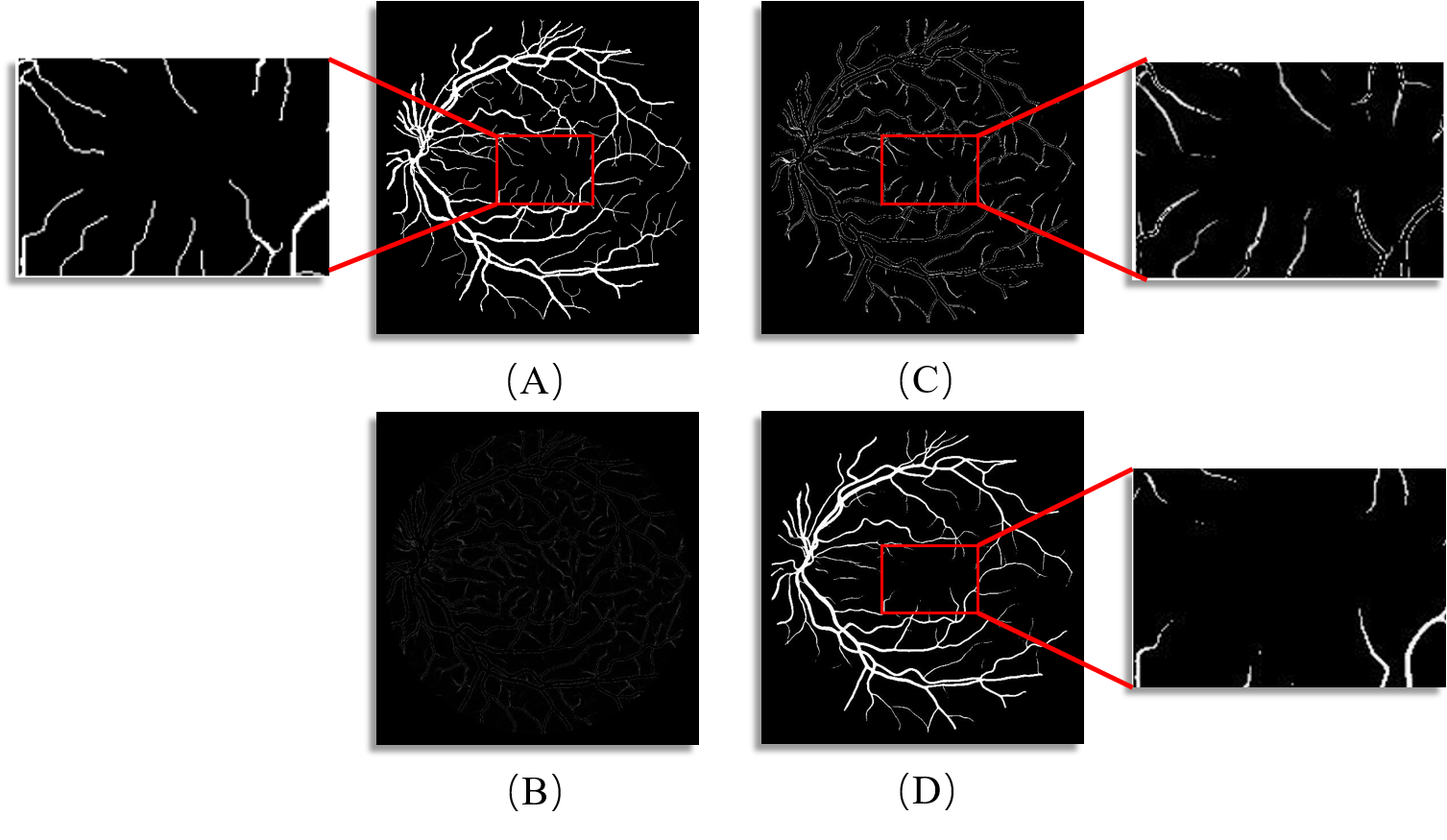}
	\caption{The visualization of probability maps of a retinal image in different ranges. (A) An ground-truth vessel segmentation from the AV-DRIVE dataset; (B) Probability maps where pixels ranging between [0,0.3]; (C) Probability maps where pixels ranging between (0.3,0.7$]$; (D) Probability maps where pixels ranging between (0.7,1.0$]$.} \label{Probability}
\end{figure}

\subsection{Spatial Activation}
Here a detailed motivation and description of the spatial activation of our capillary-enhanced network is presented. From Figure \ref{RetinalFundus}, we observe the low-contrast intensity of capillary vessels compared with the background, which leads to uncertainty from the segmentation model in classifying them. To validate this hypothesis, we extract the probability heatmap of predicting the whole vessel and visualize it in Figure \ref{Probability}. In Figure \ref{Probability} (A), the golden standard vessel segmentation is shown. In Figure \ref{Probability} (B), (C) and (D), the probability map is divided into three ranges: [0,0.3], (0.3,0,7$]$ and (0.7,1.0$]$. The logits of vessel pixels tend to 1, and those of backgrounds tend to 0. The majority of the pixels corresponding to thick vessels range between (0.7,1.0$]$. However, the pixels in the areas of capillary vessels range between (0.3,0.7$]$, indicating that the network is highly uncertain in predicting these capillary vessels, decreasing the network performance in sensitivity.

To further utilize the contextual information of vessel segmentation to promote the difficult sub-type vessel segmentation, an ad-hoc activation function is proposed, as in Equation {\ref{equ0}}:  

\begin{equation}
\label{equ0}
%\vspace{-0.4cm}
%y = \sigma e^{-(x-0.5)^{2}}+1-\sigma e^{-\frac{1}{4}} 
m(x) = \sigma (e^{-(x-\frac{1}{2})^{2}} -  e^{-\frac{1}{4}}) +1,
%\vspace{0.4cm}
\end{equation}

\begin{figure}[!t]
	\centering
	\includegraphics[width=8cm,height=4.5cm]{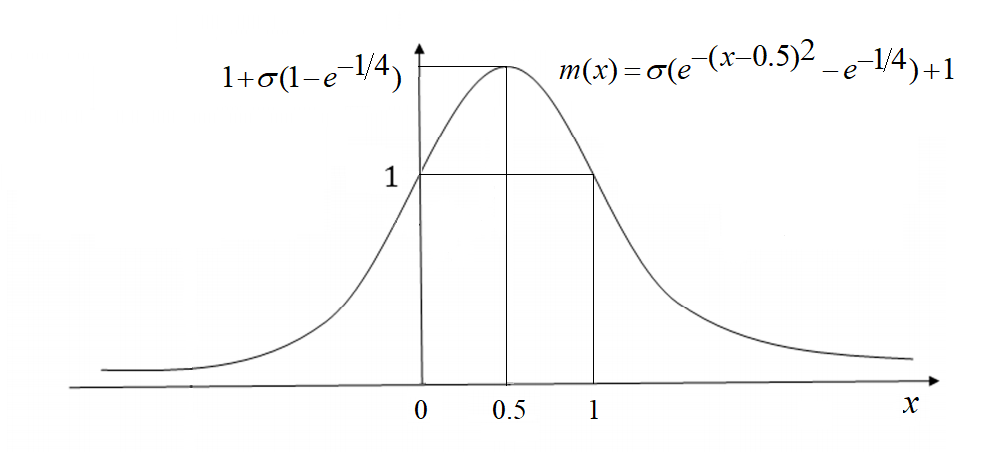}
%	\vspace{-0.3cm}
	\caption{The activation function $m(x)$ in the proposed spatial activation module.} \label{graph}
%	\vspace{-0.5cm}
\end{figure}
\begin{figure}[!t]
	\centering
	\includegraphics[width=8cm,height=3.0cm]{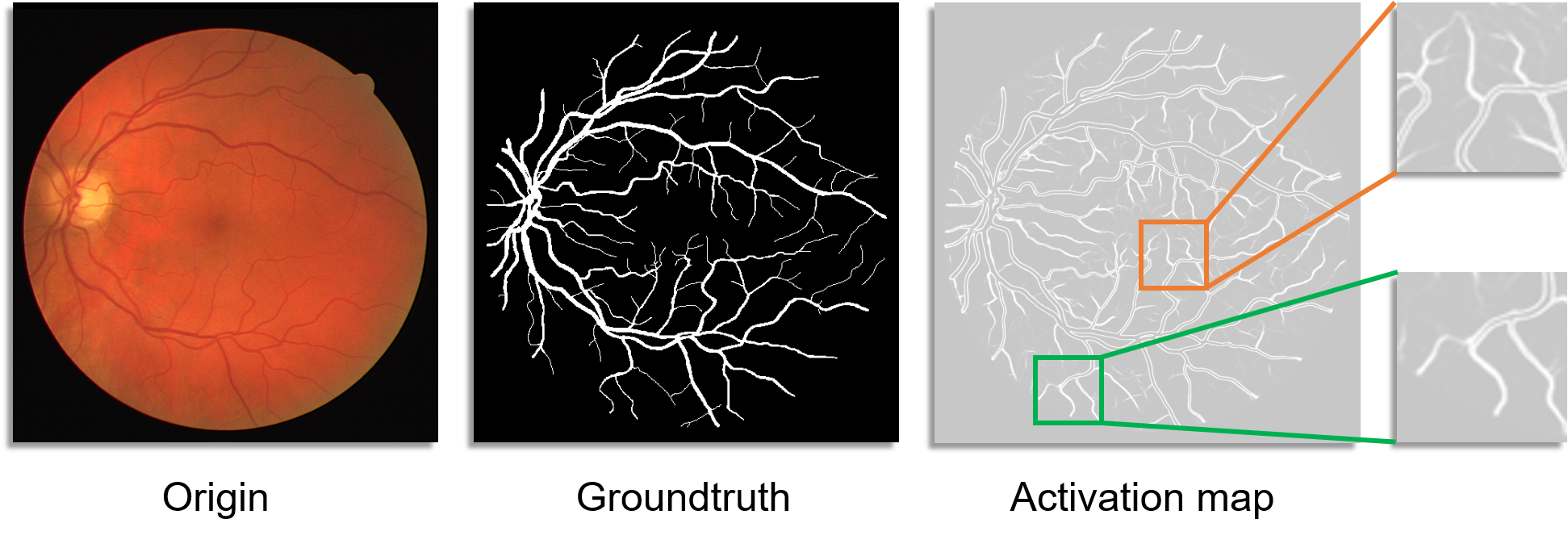}
	\caption{An example of the proposed spatial activation. The attention maps is obtained by the activation function $m(x)$ referred in Figure \ref{graph}.} \label{DrivePerformance}
\end{figure}

where $\sigma$ denotes the activation factor and is set as the identity mapping in our design, $x$ denotes the value of each pixel on the probability map with regards to whole vessel segmentation. In Figure \ref{graph} we show the graph of this activation function. It can be seen that as the value of $x$ approaches 0.5, the maximum $1+\sigma (1-e^{-\frac{1}{4}})$ of the function is approximated. As the value of $x$ approaches 0 or 1, then m(x) collapses to the identity mapping. We can pass the logit (probability map) of the whole vessel segmentation to this activation function and receive an attention weighting map. We then apply the attention score map to weight each feature maps from sub-type segmentation via point-wise multiplication.

In such a design, capillary vessels with pixel value around 0.5 will be enhanced with higher weights close to $1+\sigma (1-e^{-\frac{1}{4}})$. Meanwhile, background or thick vessels that are easier to segment will remain unchanged with weights close to 1. An example of the activation map is presented in Figure \ref{DrivePerformance}.  

A similar weighting approach is the focal loss \cite{Lin_2017_ICCV} where hard examples are assigned larger weights automatically. In our case the hard example is the accurate segmentation of the capillary vessels. However, what distinguishes our work from the focal loss is we regularize the hard examples based on another related but easier task, i.e., the vessel segmentation in an hierarchical fashion. Additionally, focal loss is only used during training while our hierarchical spatial activation is applied on both training and testing phase.

\begin{figure*}[ht]
\centering
\includegraphics[width=17.0cm,height=13.0cm]{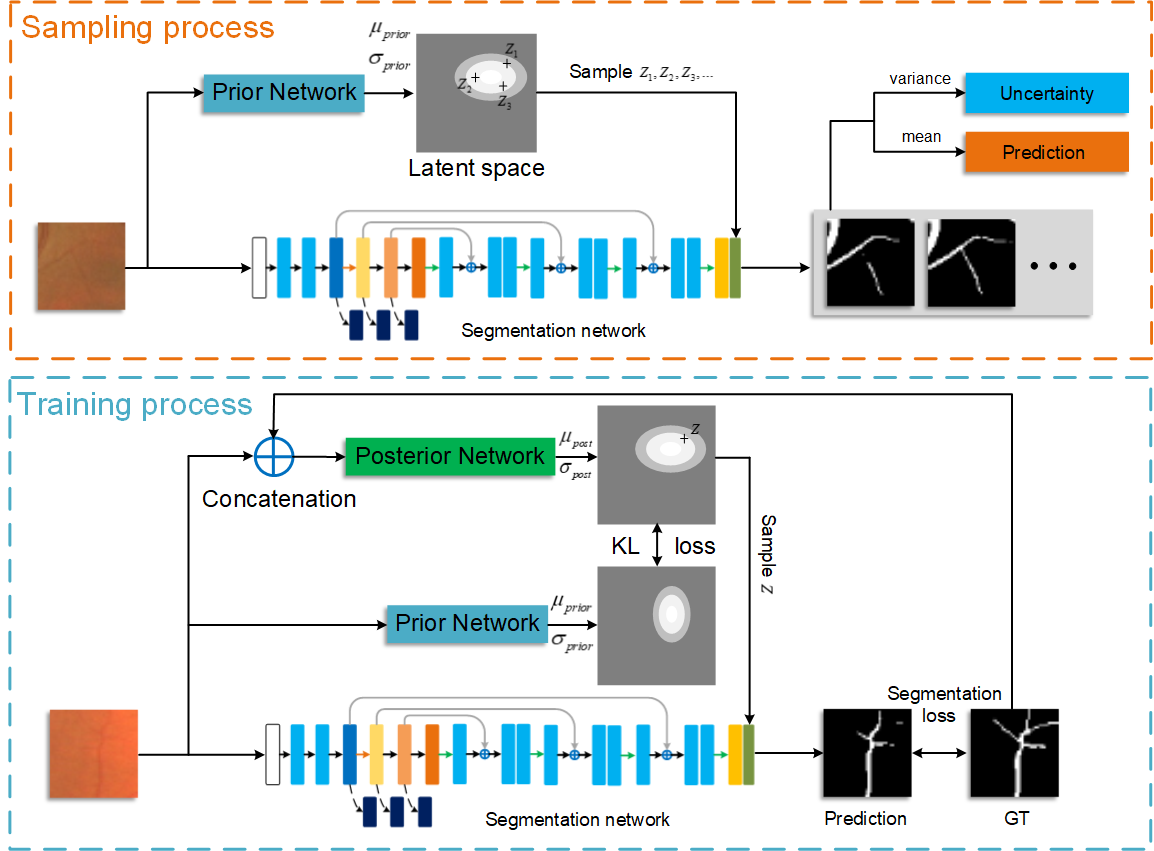}
\caption{The overall architecture of the proposed uncertainty-aware semi-supervised learning framework. The sampling process is designed to generate the pseudo annotation as well as its uncertainty estimation for an uncertainty-aware self-labeling program. The training process allows the model to capture the semantic-correlated pattern for the input unlabeled image. These two processes are alternative as well as progressive and can completes each other.
Besides the proposed hierarchical segmentation network, we introduce a prior network and posterior network to implement the above processes.
  } \label{uncertainty}
\end{figure*}

\subsection{Deep Supervision}
Lacking sufficient amount of labeled images often leads to over-fitting of a model. To deal with this problem, we adopt a deep supervision \cite{dou20163d} method, as shown in Figure \ref{fig3} (B). By adding extra branches, the shallow layers can be guided closer to supervision information, which enables low-level layers to extract more semantics \cite{zhang_exfuse:_2018}.

We design the loss function including binary cross-entropy loss of the main branch, deep supervision branches and a weight decay regularization term, as in Equation {\ref{equ1}} and {\ref{equ2}}:

\begin{equation}
\vspace{-0.2cm}
\label{equ1}
\begin{aligned}
\mathcal{L}(pred,GT) = &BCE(pred_{o},GT) \\
&+\frac{1}{3}\sum_{i=1}^{3}BCE(pred_{i},GT)+\frac{\lambda}{2} ||\Theta ||_{2}^{2},
\end{aligned}
\end{equation}

\begin{equation}
\begin{aligned}
\label{equ2}
BCE(pred,target) &= \mathcal{L}_v + \mathcal{L}_{v1} + \mathcal{L}_{v2} \\
&= -\sum_{c=1}^{3}\mu _{c}\cdot target_{c}\cdot  \log (pred_{c}),
\vspace{-0.2cm}
\end{aligned}
\end{equation}
where $\Theta$ denotes the network parameters; $pred_{o}$ denotes the output of the segmentation network; $i$ denotes the $i^{th}$ deep supervision block; $c$ denotes the $c^{th}$ class; the weight of each class is denoted as $\mu _{c}$ with $\frac{3}{7}$, $\frac{2}{7}$ and $\frac{2}{7}$ for vessel, artery and vein in retinal image and portal/hepatic vessel segmentation. The binary cross-entropy is applied on every pixel in cropped patches.

%\begin{figure*}[ht]
%\centering
%\includegraphics[width=18.0cm,height=7.5cm]{trainingProcess.png}
%\caption{The training strategy under our semi-supervised learning. (A)Supervised training %phase; (B)Unsupervised training phase.} \label{trainingProcess}
%\end{figure*}

\subsection{Uncertainty-aware Semi-supervised Segmentation}
% The model uncertainty exists in nature for learning systems, and can be used to estimate the uncertainty and quality in turn for pseudo self-labels. There are two possible sources of uncertainty \cite{der2009aleatory,kendall2017uncertainties}: epistemic uncertainty and aleatoric uncertainty. The former address the ignorance about which model generated the data while the later denote the noise inherent in the observation. 
% In medical imaging practice, uncertainty was used to display probable errors \cite{ching2018opportunities} and often aided medical image segmentation \cite{devries2018leveraging,leibig2017leveraging}. For instance, Yu \textit{et al.} \cite{yu2019uncertainty} designed a novel uncertainty-aware scheme to build a self-ensemble model for semi-supervised learning for 3D left atrium segmentation. Galdran \textit{et al.} \cite{galdran2019uncertainty} focused on pixel-level uncertainty in vessel labeling for the task of vasculature segmentation by formulating the retinal A/V classification task as a four-class segmentation problem. Since the ambiguity lies in labeling A/V vessels, probabilistic models could well parameterize the true label distribution.
Labeling vessel images is laborious and requires expert knowledge. To further leverage massive unlabeled data for better generalization, we propose a semi-supervised learning framework. Generating pseudo labels for the unlabeled data by an automatic annotation algorithm is a popular approach in semi-supervised learning. 
However, simply applying a deterministic model on the unlabeled dataset for pseudo-label acquisition ignores the common distribution shift between labeled data and unlabeled data. Once trained, the model based on labeled data may generate low-quality pseudo-labels for unlabeled data. 
Another major drawback of this labeling strategy lies in the fact that the annotation is performed by hard thresholding on the logits without handling the uncertainty in making such predictions, which leads to bias in pseudo-labels. 
% It is common in the task of vessel and sub-type vessel segmentation where the low-contrast intensities of capillaries usually cause the ambiguity in data annotation. 
To tackle these problems, we improve the pseudo-label generation by introducing the scheme of uncertainty estimation, which has been investigated in medical imaging practice \cite{devries2018leveraging,leibig2017leveraging,yu2019uncertainty,galdran2019uncertainty}. Inspired by a probabilistic U-Net for segmentation \cite{kohl2018a}, we propose an uncertainty-aware architecture based on Bayesian networks \cite{kendall2017uncertainties}, which is shown to be more efficient than Monte Carlo dropout \cite{der2009aleatory}. 

The overview of the training and sampling process of our proposed uncertainty-aware semi-supervised segmentation method is shown in Figure \ref{uncertainty}. Besides the proposed hierarchical segmentation network, we further introduce the prior network and the posterior network, to capture the prior and posterior latent representations of the input image, respectively. At a high level, the prior network is trained to embed similar representations to those by the posterior network with a Kullback-Leibler (KL) loss. Then in the sampling process, the latent representations captured by the trained prior network are expected to be close to the posterior representations and carry the label information.
Following the spirit of conditional variational autoencoder (CVAE), the latent space is modeled as an N-dimensional axis-aligned Gaussian distribution with parameters as mean $\mu$ and covariance $\sigma^2$.

More precisely, in the training process, we train the prior network with the assist of a posterior network. The posterior network is fed with the input image and its ground truth and it is trained to generate the distribution that carries the information of semantic labels, which we call as posterior distribution. Then we utilize a Kullback-Leibler (KL) divergence to constrain the prior distribution generated from the prior network to approach the posterior distribution. Once trained, the prior network is regarded to be able to capture a similar pattern to the posterior network, and the prior distribution is approximate to the posterior distribution.

Then in the sampling process, we can draw a sample $z_i$ from the approximate posterior distribution from the trained prior network. We broadcast it to an N-channel feature map and concatenate it with the output of the hierarchical segmentation network. By repeating the distribution sampling $M$ times, $M$ corresponding model outputs can be obtained. Then the pixel-wise mean values and variance values among them are calculated as the pseudo annotation and the uncertainty estimation. Please note that for an input image, the output of the segmentor (hierarchical capillary-enhanced network) is invariant during different samples of latent space, so the computation of these forward passes is cheap.

With the pseudo-labels with uncertainty estimation, we can improve the hierarchical capillary-enhanced model by retraining only the pseudo-labels with high confidence. Assume that the model uncertainty of input image is $U$, which is expressed as the variance computed by several forward passes of the sampling. The notation $u_v$ represents the uncertainty at the $v$-th pixel in the pseudo-label. The Loss function of the retraining can be formulated in Equation {\ref{equ_unsupervised}} as:

\begin{equation}
\label{equ_unsupervised}
\mathcal{L}_{u}(pred,pseudo)=\frac{\sum_{v}^{}(\mathbb{I}(u_v<H)\cdot BCE(pred,pseudo))}{\sum_{v}^{}\mathbb{I}(u_v<H)},
\vspace{-0.2cm}
\end{equation}
where $\mathbb{I}(\cdot )$ is the indicator function outputting 1 if the condition is satisfied and 0 otherwise; $pred$ and $pseudo$ are the segmentation prediction and the pseudo label at the $v$-th pixel; $H$ is a threshold to filter out uncertain label empirically set as 0.7. With this uncertainty-aware loss function, the segmentation network can learn from the reliable pseudo labels as well as avoid the possible misleading from unreliable ones. The overall algorithm for the proposed semi-supervised learning framework is list in Algorithm 1.

\begin{algorithm}[htb]  
  \caption{Overall flow of the proposed semi-supervised learning framework}  
  \label{alg:Framwork}  
  \begin{algorithmic}[1]  
    \Require  
      labeled dataset $\mathcal{D}_l$, unlabeled dataset $\mathcal{D}_u$; pretraining iterations $T_{pre}$, retraining iterations $T_{re}$;
      initialized parameter of segmentation network $\theta$, initialized parameter of prior and posterior networks $\theta^{'}$.
    \Ensure  
      trained parameter $\theta$ of segmentation network. 

    \For{$loop$ = 1 to $T_{pre}$ }  
    \State Optimize $\theta$ and $\theta^{'}$ on labeled dataset $\mathcal{D}_l$ by the training process in Figure \ref{uncertainty}.
    \EndFor  
    \For{$loop$ = 1 to $T_{re}$ }  
      \State Take multiple forward passes of pretrained model to generate outputs on unlabeled dataset $\mathcal{D}_u$ by the sampling process in Figure \ref{uncertainty}.
      \State Calculate the mean values of the multiple outputs as pseudo annotation.
      \State Calculate the variance values of the multiple outputs as uncertainty estimation.
      \State Optimize $\theta$ on dataset $\mathcal{D}_u$ with the pseudo annotation and uncertainty estimation using Equation \ref{equ_unsupervised};
    \EndFor  
\end{algorithmic}  
\end{algorithm}  

\section{Experimental Results}
\subsection{Dataset}
\begin{figure}[!t]
	\centering
	\includegraphics[width = 8cm]{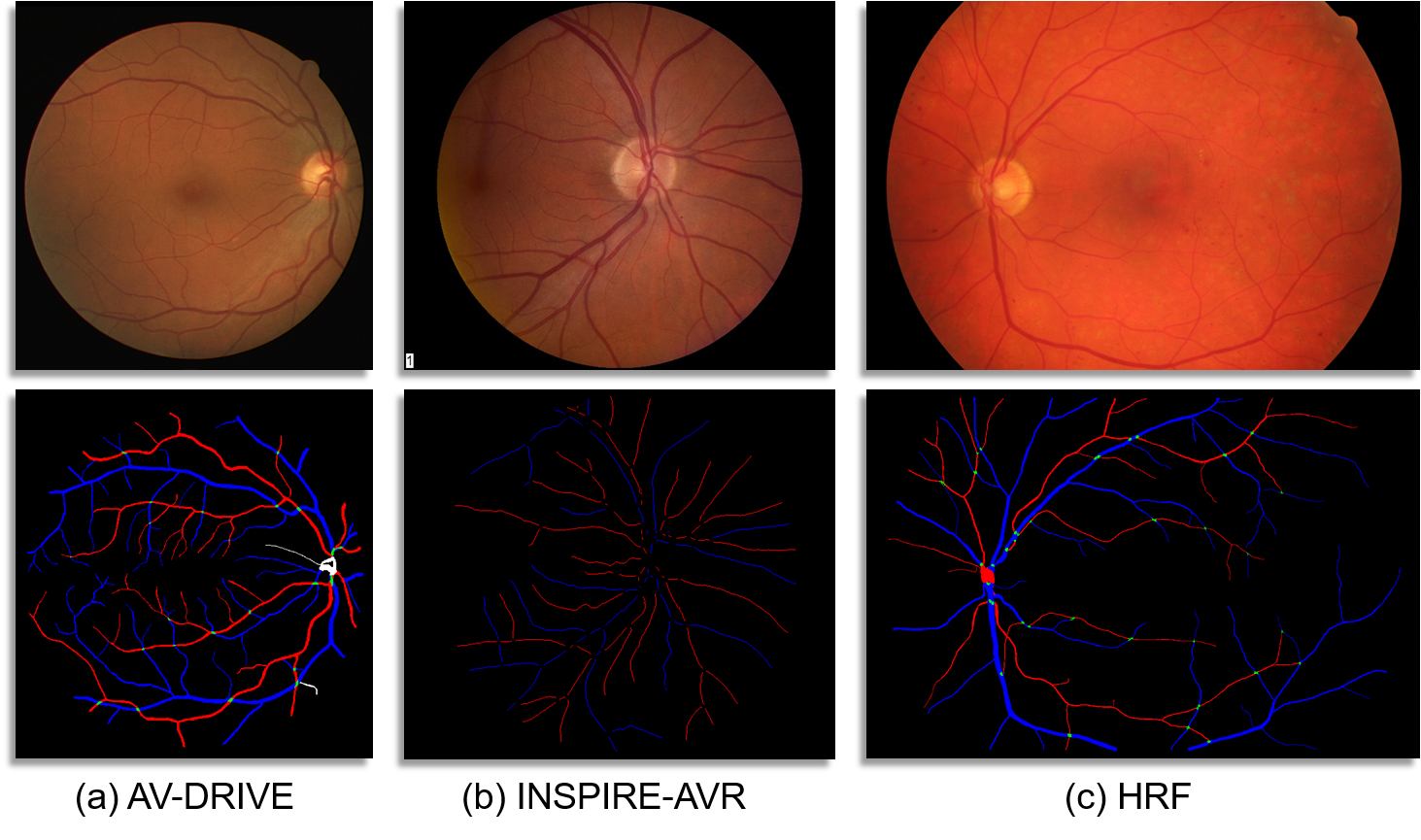}
	\caption{Some examples of the public database (first row) and their corresponding manual segmentation (second row). Please note, there are only centerline-annotated labels in INSPIRE-AVR, which is different from AV-DRIVE and HRF. } 
	\label{dataset}
\end{figure}

For retinal artery/vein segmentation, we trained and evaluated our model on the publicly available AV-DRIVE database \cite{Hu2013_drive}. The AV-DRIVE database contains 20 training and 20 testing colorful retinal fundus images with dimension of $584\times565$ pixels, with pixel-wise labeling of vessel segmentation and A/V classification provided. To validate our semi-supervised learning approach, we pretrained segmentation model on AV-DRIVE dataset and generate pseudo annotations on HRF \cite{odstrcilik2013retinal} dataset which contains 45 images of the dimension 800 $\times$ 1200. We further retrained our model on the combination of AV-DRIVE and HRF datasets and tested it on the INSPIRE-AVR dataset containing 40 fundus images with dimension of $2048\times2392$. Some examples of the public database can be seen in Figure \ref{dataset}.

For the liver portal/hepatic vessel segmentation, we adopted the IRCAD datasets containing 3D CT-scans of 20 different imaging subjects. All the 20 scans are labeled with the pixels of portal and hepatic vessel. We conducted K-fold cross-evaluation and report the mean results of them following prior study \cite{lebre2019automatic,merveille2017curvilinear,huang2018robust,jerman2016enhancement,guo2020novel}. 

\subsection{Implementation Details}
In the training stage, the stochastic gradient descent algorithm with momentum was adopted to optimize our model. The initial learning rate was set as 0.05 and halved every 5000 iterations. The size of the training mini-batch was 16. Multi-scale patches with size of $64\times64, 96\times96, 128\times128$ were randomly cropped from the retinal images during training and were further resized to $64\times64$ to exploit multi-scale information. In the testing stage, patches with size of $64\times64$ were extracted at the stride of 10. The segmentation of sub-type vessels and full vessel of a image were obtained by aggregating the patches, with the overlapping regions being averaged as post-processing. The experiments were conducted on a single NVIDIA GPU (Tesla P40) and the training process took about 3 hours. 

\subsection{Evaluation Metrics}
We adopt common evaluation metrics: Accuracy(Acc), Sensitivity(Sen), Specificity(Sp), Area Under Curve(AUC). The Accuracy, Sensitivity and Specificity are defined as

\begin{equation}
\label{equ3}
Accuracy = \frac{TP+TN}{TP+TN+FP+FN},
%\vspace{-0.2cm}
\end{equation}

\begin{equation}
\label{equ4}
Sensitivity = \frac{TP}{TP+FN},
%\vspace{-0.2cm}
\end{equation}

\begin{equation}
\label{equ4}
Specificity = \frac{TN}{FP+TN},
%\vspace{-0.2cm}
\end{equation}

The metric AUC is defined as the receiver operating characteristics (ROC) area under the curve, which represents the probability that a classifier will rank a randomly chosen positive instance higher than a randomly chosen negative one. 
%Dice Coefficient is defined as follow
%\begin{equation}
%\label{equ5}
%Dice(S,L) = \frac{2|S\cap L|}{|S|+|L|}\times100\%,
%\vspace{-0.2cm}
%\end{equation}
 %The upper view shows a case of complex crossovers, where a vessel crosses another near branch point, posing a major challenge for the graph based method. The lower view shows a case of two close-by parallel vessels with a thin gap, which might confuse the vessel segmentation algorithm by taking them as one single thick vessel. 
%where S denotes the set of segmentation prediction and L denotes the set of segmentation label. The Dice Coefficient ranges 0 to 1 and measures the overlap between the segmentation prediction and segmentation label.

\begin{table*}[!t]
	\centering
	\caption{The ablation study results of vessel segmentation and A/V classification.}\label{abstudy}
	\vspace{-0.2cm}
	\begin{tabular}{p{0.8cm}<{\centering}p{0.6cm}<{\centering}p{0.6cm}<{\centering}|p{1.3cm}<{\centering}|p{1.3cm}<{\centering}|p{1.3cm}<{\centering}|p{1.3cm}<{\centering}|p{1.3cm}<{\centering}|p{1.3cm}<{\centering}|p{1.3cm}<{\centering}}
		\hline
		\multicolumn{3}{c|}{Combination}&\multicolumn{4}{c|}{Vessel Segmentation}&\multicolumn{3}{c}{A/V Classification}\\
		%& \multicolumn{|4|4} {& AV-DRIVE &  HRF}\\
		\hline
		MTs & MIs & AC & Acc($\%$) & Sen($\%$) & Sp($\%$) & AUC($\%$)  & Acc($\%$) & Sen($\%$) & Sp($\%$) \\ \hline
		&  & & 94.98  & 68.86 & \textbf{98.79} & 97.60  &  91.25 & 89.68 & 92.55 \\ %\hline
		%\checkmark & & &   & 0.9499  & 0.6887  &  \textbf{0.9880}  & 0.9770  & 0.9137 & 0.8862 & \textbf{0.9360} \\ \hline
		
		 \checkmark &  &  & 95.61  & 78.50  &  98.10  & 98.01  & 91.63  & 90.46  & 92.63  \\ %\hline
		
		 \checkmark &\checkmark  & & 95.66 & 78.30 & 98.19 & 98.08 & 91.98  & 90.36  & \textbf{93.42}   \\
		%\hline
		%		\checkmark & \checkmark &\checkmark &\checkmark   & 0.9565 & 0.7891 & 0.9811 & 0.9808 & 0.9168  & 0.9079  &0.9230   \\
		%		\hline
		 \checkmark &\checkmark  &\checkmark  & \textbf{95.70} & \textbf{79.16} & 98.11 & \textbf{98.10}  & \textbf{92.58}& \textbf{92.18} & 92.98 \\ \hline
	\end{tabular}
%\vspace{-0.3cm}
\end{table*}

\begin{table*}[ht]
  \centering
  \caption{Performance comparison of A/V segmentation of semi-supervised learning.}\label{tab_uncertainty}
  \begin{tabular}{c| p{1.5cm}<{\centering} |p{1.5cm}<{\centering} |p{1.5cm}<{\centering}}
  %|p{1.5cm}<{\centering} |p{2.0cm}<{\centering}}
  \hline
  Methods & Acc($\%$) & Sp($\%$) & Sen($\%$) \\
  %\hline
  %Fraz et.al \cite{fraz_ensemble_2012} (2012)&  0.9480 & 0.9807 & 0.7406 & 0.9747\\
  \hline
  Supervised & 84.7 & 92.5 & 77.1 \\
  %\hline
  Co-training & 91.6 & 92.4 & 91.3  \\
  %\hline
  %Li \textit{et al.} \cite{li_cross-modality_2016} (2016)& 0.9527 & 0.9816 & 0.7569 & 0.9738 \\ \hline
  Semi-supervised (Vanilla Annotation) & 85.4 & 95.8 & 64.5   \\
  
  Semi-supervised (MC Dropout) & 88.3 & 88.9 & 88.2   \\
  %\hline
  %Dasgupta et.al \cite{dasgupta_fully_2017} (2017& 0.9533 & 0.9801 & 0.7691 & 0.9744 \\
  %\hline
  \bfseries Semi-supervised (Proposed) & 90.1  & 88.2 & 92.4   \\
  \hline
  \end{tabular}
  \end{table*}

\subsection{Ablation Studies}
On the retinal fundus image dataset AV-DRIVE, three groups of ablation studies are performed to evaluate the contribution of each design in our approach: 1) Joint prediction of vessel segmentation and artery/vein segmentation in a multi-task manner (denoted as MTs); 2) Multiple inputs using pre-processed images (denoted as MIs); 3) Spatial activation (denoted as AC). Baseline is built by removing the MTs, MIs and AC designs.

\begin{figure}[!t]
	\centering
	\includegraphics[width = 8cm]{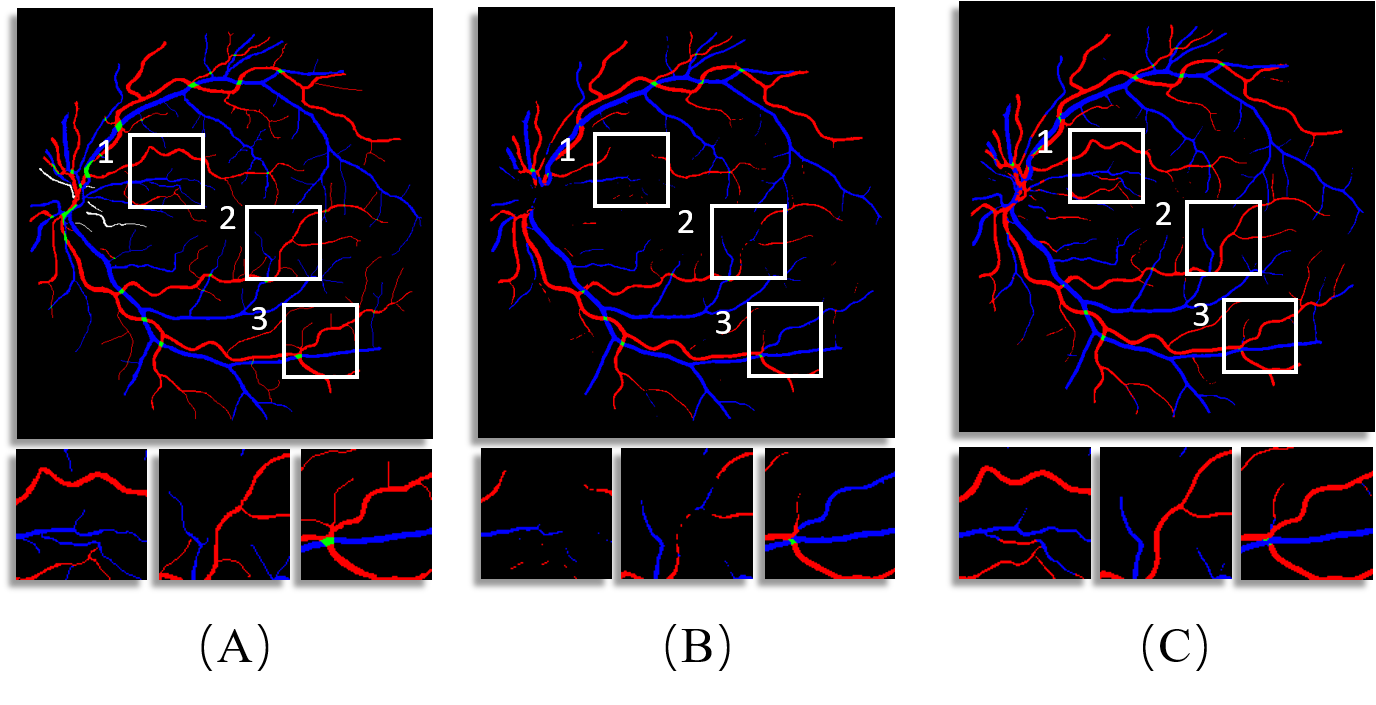}
	\caption{Comparison of model performance for baseline and proposed method. (A) Ground truth; (B) Baseline model performance; (C) Proposed method.} 
	\label{AbstudyRep}
\end{figure}

As indicated in Table \ref{abstudy}, a multi-task way of performing vessel segmentation and sub-type segmentation simultaneously can improve both tasks. Also, by inputting multiple pre-processed images into the model, the A/V classification accuracy is improved by 0.3\%. When the spatial activation is applied, the A/V segmentation can be further increased by 0.6\%. Our proposed hierarchical capillary-enhanced network achieves a pixel-wise accuracy of 95.70\% for vessel segmentation and 92.58\% for sub-type vessel segmentation. We also compare our spatial activation with the design of focal loss. We observe no significant improvement when removing the activation block and equipping the cross-entropy loss with focal loss. In Figure \ref{AbstudyRep}, an example is shown to compare our model with baseline. We zoom in multiple local areas, and our proposed method has remarkably improved the segmentation result of retinal artery and vein over the baseline by the MTs, MIs and AC design.

\begin{table*}[ht]
  \centering
  \caption{Performance comparison of whole vessel segmentation on the AV-DRIVE dataset.}\label{tab2}
  \begin{tabular}{c| p{1.5cm}<{\centering} |p{1.5cm}<{\centering} |p{1.5cm}<{\centering}| p{1.5cm}<{\centering}}
  \hline
  Methods & Acc($\%$) & Sp($\%$) & Sen($\%$) & AUC($\%$)\\
  %\hline
  %Fraz et.al \cite{fraz_ensemble_2012} (2012)&  0.9480 & 0.9807 & 0.7406 & 0.9747\\
  \hline
  Fu \textit{et al.} \cite{fu2016retinal} & 94.70 & - & 72.94 & - \\
  %\hline
  Liskowski \textit{et al.} \cite{liskowski_segmenting_2016} & 95.35 & 98.07 & 78.11 & 97.90 \\
  %\hline
  %Li \textit{et al.} \cite{li_cross-modality_2016} (2016)& 0.9527 & 0.9816 & 0.7569 & 0.9738 \\ \hline
  Mo \textit{et al.} \cite{mo_multi-level_2017} & 95.21 & 97.80 & 77.79 & 97.82 \\
  %\hline
  %Dasgupta et.al \cite{dasgupta_fully_2017} (2017& 0.9533 & 0.9801 & 0.7691 & 0.9744 \\
  %\hline
  Wu \textit{et al.} \cite{wu_multiscale_2018} & 95.67 & \textbf{98.19} & 78.44 & 98.07 \\
  %\hline
  \bfseries Proposed   & \textbf{95.70} & 98.11 & \textbf{79.16} & \textbf{98.10} \\
  \hline
  \end{tabular}
  \end{table*}
  
  \begin{table*}[ht]
    \centering
    \caption{Performance comparison of A/V classification on AV-DRIVE and INSPIRE datasets.}\label{tab3}
    \begin{tabular}{c|p{1.2cm}<{\centering}|p{1.2cm}<{\centering}|p{1.2cm}<{\centering}|p{1.2cm}<{\centering}|p{1.2cm}<{\centering}|p{1.2cm}<{\centering}}
    \hline
    \multicolumn{1}{c|}{}&\multicolumn{3}{c|}{AV-DRIVE}&\multicolumn{3}{c}{INSPIRE}\\
     %\cmidrule(r){2-4}  \cmidrule(r){5-7}
     \cline{2-7}
    Methods & Acc($\%$) & Sen($\%$) & Sp($\%$) & Acc($\%$) & Sen($\%$) & Sp($\%$) \\
    \hline
    Dashtbozorg \textit{et al.} \cite{dashtbozorg_automatic_2014}  & 87.4 & 90.0 & 84.0 &84.9 &91.0 &86.0 \\ %\hline
    Estrada \textit{et al.} \cite{estrada_retinal_2015} & 93.5 & 93.0 & 94.1 &90.9 &91.5 &90.2 \\ %\hline
    Xu \textit{et al.} \cite{xu_improved_2017}  & 92.3 & 92.9 & 91.5 &- &- &- \\ %\hline
    Zhao \textit{et al.} \cite{zhao_retinal_2018}  & - &91.9  & 91.5 &91.0 &91.8 &90.2\\ %\hline
    %\bfseries Proposed (GT)  & 92.6 & 92.2 & 93.0 & 90.3&91.4 &89.7\\ %\hline
    \bfseries Proposed & \textbf{94.5} & \textbf{93.4} & \textbf{95.5} &\textbf{91.6} &\textbf{92.4} &\textbf{91.3} \\ \hline
    
    \end{tabular}
    \end{table*}
    
To evaluate our semi-supervised learning framework, we pretrain a segmentation model using the labeled AV-DRIVE dataset and generate pseudo labels on the unlabeled HRF dataset.
The evaluation is conducted on the INSPIRE database, denoted as Semi-supervised (Proposed). 
We first evaluate the benefits of additional datasets using a trained segmentor on AV-DRIVE to directly annotate images on the HRF dataset without uncertainty modeling, denoted as Semi-supervised (Vanilla Annotation).
In order to validate our uncertainty estimation approach, we conduct a comparative experiment to replace it with the Monte Carlo Dropout \cite{gal2016dropout}, denoted as Semi-supervised (MC Dropout).
Since the manual segmentation labels of HRF dataset are available, we build the performance upper bound by training the proposed hierarchical network by both manually well-labeled AV-DRIVE and HRF datasets and testing it on INSPIRE dataset, denoted as Co-training.
As a baseline, a segmentation network is trained only on AV-DRIVE database and test on INSPIRE, denoted as supervised. 

The comparative experimental results on A/V segmentation can be seen in Table \ref{tab_uncertainty}. The Co-training model achieves better segmentation than baseline Supervised model, due to the augmented data size. Besides, when applying our uncertainty estimation model, semi-supervised method produces more accurate segmentation than the one using MC Dropout, approaching the upper bound Co-training. The vanilla annotation in such a semi-supervised framework brings little improvements, because it introduces massive unreliable pseudo labels where the model is misled.

\begin{figure}[h]
	\centering
	\includegraphics[width = 8cm, height=14cm]{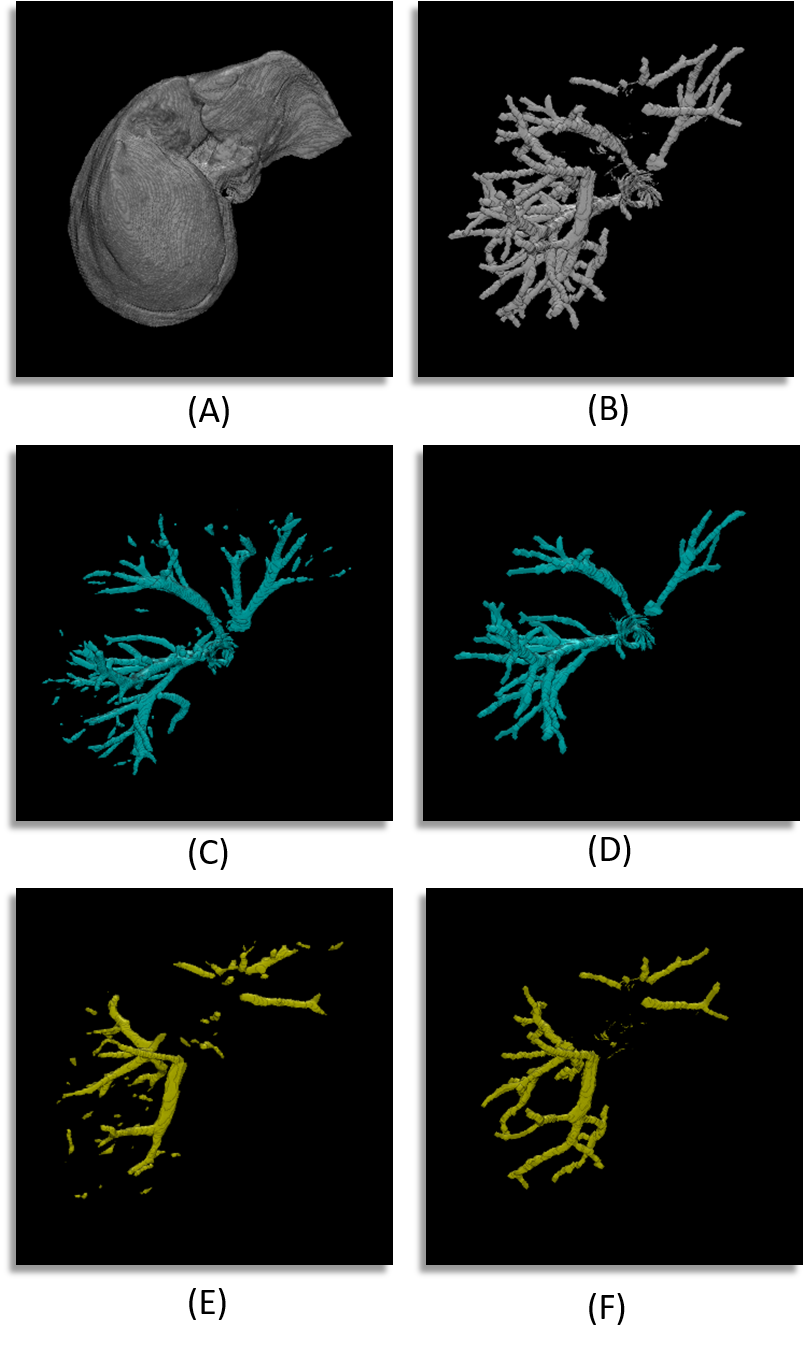}
	\caption{The visualization of segmentation of portal vessel and hepatic vessel. (A) CT liver volume from IRCAD dataset; (B) manual label of portal and hepatic vessel; (C) segmentation prediction of hepatic vessel; (D) manual label of hepatic vessel; (E) segmentation prediction of portal vessel; (F) manual label of portal vessel.} 
	\label{liver_image}
\end{figure}

\begin{table*}[ht]
  \centering
  \caption{Performance comparison of hepatic and portal vessel segmentation on the IRCAD dataset.}\label{tab_liver}
  \begin{tabular}{c|p{1.2cm}<{\centering}|p{1.2cm}<{\centering}|p{1.2cm}<{\centering}|p{1.2cm}<{\centering}|p{1.2cm}<{\centering}|p{1.2cm}<{\centering}|p{1.2cm}<{\centering}|p{1.2cm}<{\centering}|p{1.2cm}<{\centering}}
  \hline
  \multicolumn{1}{c|}{}&\multicolumn{3}{c|}{Vessel Segmentation}&\multicolumn{3}{c|}{Portal Vessel Segmentation} &\multicolumn{3}{c}{Hepatic Vessel Segmentation}\\
   %\cmidrule(r){2-4}  \cmidrule(r){5-7}
   \cline{2-10}
  Methods & Acc($\%$) & Sen($\%$) & Sp($\%$) & Acc($\%$) & Sen($\%$) & Sp($\%$) & Acc($\%$) & Sen($\%$) & Sp($\%$) \\
  \hline 
  Merveille \textit{et al.} \cite{merveille2017curvilinear} & 90.0 & 20.0 & 97.0 &98.0 &56.0&\textbf{99.0}&96.0&23.0&98.0 \\
  Lebre \textit{et al.} \cite{lebre2019automatic}  & 97.0 & 69.0 & 98.0 & 97.0 & \textbf{70.0} & 98.0 & 98.0 & 54.0 & 98.0 \\ %\hline
  Huang \textit{et al.} \cite{huang2018robust} & 97.1 & 74.3 & 98.3 &-&-&-&-&-&- \\ 
  Jerman \textit{et al.} \cite{jerman2016enhancement} & 97.2 & \textbf{76.7} & 96.5 &-&-&-&-&-&-  \\
  Guo \textit{et al.} \cite{guo2020novel} & 97.8 & 66.2 & \textbf{98.7} &-&-&-&-&-&-  \\
    
  \bfseries Proposed  & \textbf{97.9} & 76.4 & 98.6 & \textbf{98.9} & 61.5 & 98.3 & \textbf{99.7}& \textbf{62.4} & \textbf{99.4} \\ \hline
  \end{tabular}
  \end{table*}

\subsection{Comparison to Existing Methods}
We compare our hierarchical capillary-enhanced network with other state-of-the-art approaches in the performances of whole retinal fundus vessel and artery/vein segmentation, as shown in Table \ref{tab2} and Table \ref{tab3}.  

In Table \ref{tab2}, we compare the capillary-enhanced network with other state-of-the-art models for whole retinal vessel segmentation on the testing dataset in AV-DRIVE. Our approach achieves the better performance among the compared models, owing to the joint optimization for both whole and sub-type vessel segmentation. 

In Table \ref{tab3}, we compare our capillary-enhanced network with other state-of-the-art models for retinal arteries/veins segmentation on the testing dataset in both AV-DRIVE and INSPIRE datasets.
% Conditioned on the fact that existing methods focused on the sub-type vessel segmentation accuracy only based on accurately segmented vessels, which overlooks the negative effect of inaccurate segmentation on whole vessel, we use a more strict but practical evaluation criteria to account for the inaccuracies in segmenting whole vessels. 
When performing under a fair comparison,  the proposed model achieves a pixel-wise accuracy of 94.50\%, which outperforms the current state-of-the-art A/V segmentation method by a considerable margin.

%In order to further prove the performance of proposed model, we have evaluated the A/V classification performance on the INSPIRE dataset. The training set contains 20 images from AV-DRIVE and 45 images from HRF dataset, which contains publicly available vessel segmentation label and we labeled A/V class. Our model outperforms the existing A/V classification approaches, which can achieve 91.6\% without fine-tuning.

We also have validated our algorithm on liver vessel segmentation in IRCAD \cite{IRCAD}. Because 3D liver Computed Tomography (CT) images are provided in this dataset, we adopt a 3D U-Net, incorporating the proposed designed spatial activation and removing our multi-input module. In this database, we use our model to segment the portal and hepatic vessels in a multi-task manner. Then we combine them to form the segmentation of whole veins in a liver. The 3D reconstructions of the segmented portal and hepatic vessels are shown in Figure \ref{liver_image}. 
We have compared our model with extensive prior approaches, as shown in Table \ref{tab_liver}.
Note that we only report performances on whole vessel segmentation of some compared methods \cite{huang2018robust,jerman2016enhancement,guo2020novel}, because they were proposed for targeting on whole vessel regions.
At last, the proposed method achieves superior performance against other state-of-the-art methods. The common improvements in the joint task and the two sub-tasks also demonstrate the extensive effectiveness of the proposed multi-task mechanism

% \begin{table*}[ht]
% \centering
% \caption{Performance comparison of hepatic and portal vessel segmentation on the IRCAD dataset.}\label{tab_liver}
% \begin{tabular}{c|p{1.2cm}<{\centering}|p{1.2cm}<{\centering}|p{1.2cm}<{\centering}|p{1.2cm}<{\centering}|p{1.2cm}<{\centering}|p{1.2cm}<{\centering}|p{1.2cm}<{\centering}|p{1.2cm}<{\centering}|p{1.2cm}<{\centering}}
% \hline
% \multicolumn{1}{c|}{}&\multicolumn{3}{c|}{Vessel Segmentation}&\multicolumn{3}{c|}{portal vessel Segmentation} &\multicolumn{3}{c}{hepatic vessel Segmentation}\\
%  %\cmidrule(r){2-4}  \cmidrule(r){5-7}
%  \cline{2-10}
% Methods & Acc($\%$) & Sen($\%$) & Sp($\%$) & Acc($\%$) & Sen($\%$) & Sp($\%$) & Acc($\%$) & Sen($\%$) & Sp($\%$) \\
% \hline

% Lebre \textit{et al.} \cite{lebre2019automatic}  & 97.0 & 69.0 & 98.0 & - & - & - & - & - & - \\ %\hline
%  %\hline
% \bfseries Proposed  & 97.4 & 69.4 & 98.8 & 98.9 & 52.5 & 98.1 & 99.7 & 62.4.0 & 99.4 \\ \hline
% \end{tabular}
% \end{table*}

\section{Conclusion}
In this paper, we build the learning model that faces several serious practical challenges in organ vessel segmentation. 
First of all, we propose the hierarchical capillary-enhanced network to enhance the performance of vessel segmentation, especially for capillary regions.
The proposed method enables the end-to-end segmentation of whole vessels and their sub-types simultaneously in a multi-task manner and allows the sub-tasks can benefit from the joint task. 
Moreover, we develop an uncertainty-aware semi-supervised learning framework to alleviate the practical issue of annotation scarcity.
We evaluate the proposed overall method on some publicly available databases and compare it to the prior methods extensively in both the retinal fundus datasets and liver CT set.
The experimental results show that our hierarchical capillary-enhanced model outperforms the state-of-the-art supervised methods under the fully annotated setting and other semi-supervised methods under the annotation-scarcity setting, which demonstrates the proposed model is effective extensively in the medical practice.

To summary, the proposed method significantly improves the accuracy and efficiency of A/V segmentation as well as alleviates the rely upon vessel annotation, which are both key requirements in real-world clinical applications.
In the future work, we will go on working based on this research and develop a fully automatic and annotation-efficient machine for vascular parameter modeling to facilitate the clinical vascular biomarker study.

\section*{Compliance with ethical standards}
{\bfseries Conflict of interest} This work is supported in part by National Key Research and Development Program of China (No. 2019YFC0118100), in part of ZheJiang Province Key Research Development Program (No. 2020-C03073), in part by National Natural Science Foundation of China under Grants 81671766, 61971369, U19B2031, U1605252, 61671309, in part by Open Fund of Science and Technology on Automatic Target Recognition Laboratory 6142503190202, in part by Fundamental Research Funds for the Central Universities 20720180059, 20720190116, 20720200003, and in part by Tencent Open Fund.

\footnotesize
\bibliographystyle{splncs04}
\bibliography{MIMT-Net}

% that's all folks
\end{document}